\documentclass[12pt]{article}
\usepackage{epsfig}
\usepackage{amssymb}

\def\be{\begin{equation}}
\def\ee{\end{equation}}
\def\ba{\begin{array}{c}}
\def\ea{\end{array}}

\def\ben{$$}
\def\een{$$}

\newcommand{\pkt}{\!\!\succ\,\,}
\newcommand{\kt}{\rangle}

\def\bea{\begin{eqnarray}}
\def\eea{\end{eqnarray}}

\def\beax{\begin{eqnarray*}}
\def\eeax{\end{eqnarray*}}

\begin{document}


\vspace{.35cm}

\begin{center}

{\Large

 {\em Ad hoc} physical Hilbert spaces in Quantum Mechanics

}
\end{center}

\vspace{10mm}

\begin{center}
\textbf{Francisco M. Fern\'{a}ndez and Javier Garcia}

\vspace{3mm}

INIFTA (UNLP, CCT La Plata-CONICET), Divisi\'{o}n Qu\'{\i}mica
Te\'{o}rica, Blvd. 113 S/N, Sucursal 4, Casilla de Correo 16, 1900
La Plata, Argentina

\vspace{3mm}

{e-mail:fernande@quimica.unlp.edu.ar}

\vspace{3mm}

and

\vspace{3mm}

\textbf{Iveta Semor\'{a}dov\'{a} and Miloslav Znojil}

\vspace{3mm} Nuclear Physics Institute ASCR,

250 68 \v{R}e\v{z}, Czech Republic

{e-mail: znojil@ujf.cas.cz}


\end{center}



\section*{Abstract}

The overall principles of what is now widely known as PT-symmetric
quantum mechanics are listed, explained and illustrated via a few
examples. In particular, models based on an elementary local
interaction V(x) are discussed as motivated by the naturally
emergent possibility of an efficient regularization of an otherwise
unacceptable presence of a strongly singular repulsive core in the
origin. The emphasis is put on the constructive aspects of the
models. Besides the overall outline of the formalism we show how the
low-lying energies of bound states may be found in closed form in
certain dynamical regimes. Finally, once these energies are found
real we explain that in spite of a manifest non-Hermiticity of the
Hamiltonian the time-evolution of the system becomes unitary in a
properly amended physical Hilbert space.

%
%

\noindent

\newpage

\section{Introduction\label{intr}}

\subsection{The methodical framework}

The scope and range of any study of mathematical structures as used
in quantum theory strongly depend on the domain of prospective
applications. In this sense we feel inspired, in particular, by the
success of nuclear physicist's strategy of the so called interacting
boson models as reviewed, say, in Ref.~\cite{Geyer}. This review may
be recalled as illustrating the efficiency of a judicious,
nontrivial choice of the Hilbert space, due to which the states of a
given quantum system may be represented in an optimal,
computation-friendly manner.

In the concrete nuclear-physics phenomenological setting as
presented in Ref. \cite{Geyer} the argument supporting the
importance of availability of several alternative Hilbert spaces
${\cal H}$ yielding equivalent physical predictions may be made more
specific. Indeed, it is well known that whenever we perceive an
atomic nucleus in terms of its nucleonic (i.e., fermionic) degrees
of freedom, it is just a routine task to recall the principle of
classical-quantum correspondence and to postulate the existence of a
``realistic'' Hamiltonian in its widely accepted kinetic- plus
potential-energy ``microscopic'' form of an operator
$\mathfrak{h}=\mathfrak{h}^{(A)} =-\sum^A_i \triangle_i + \sum_j
\sum_k V_{jk}$. This operator is defined and, by construction,
safely self-adjoint in the traditional Hilbert space ${\cal H}^{(P)}
\equiv L^2(\mathbb{R}^{3A})$ (here, using the notation as introduced
in our review paper \cite{SIGMA}, the superscript $^{(P)}$ stands
for ``primary'' {\it alias} ``physical'').

In the framework of nuclear physics, unfortunately, the {\em
practical\ } success of the use and study of the realistic
Hamiltonians $\mathfrak{h}^{(A)}$ remained restricted just to the
very light nuclei. For any heavier (and, in particular, heavy)
atomic nucleus, the ``brute-force'' numerical diagonalization
technique failed to provide a satisfactory precision of energies (in
numerical context) and/or a satisfactory intuitive insight into the
structure of wave functions (say, in the context of testing the
predictions experimentally).

In a way reviewed in \cite{Geyer}, a decisive and persuasive
progress has been achieved after a replacement of the microscopic,
fermionic Hamiltonians by their various effective (and, in principle
at least, isospectral) partners $H=H^{(A)}=\Omega^{-1}
\mathfrak{h}^{(A)}\,\Omega\, $. A particularly productive principle
of construction of the latter effective Hamiltonians $H^{(A)}$
relied upon the intuitively appealing idea that due to certain
specific (and more or less known) features of the ``realistic''
inter-nucleon forces $V_{jk}$, the pairs of fermions inside a
nucleus may often be perceived as coupled, intuitively at least,
into certain effective, bosons-resembling quasi-particles.

In the purely pragmatic context, an exceptional and by far the most
successful implementation of the latter constructive recipe has been
found in an appropriate adaptation of the Dyson's older idea
\cite{Dyson} by which one makes an ansatz $\mathfrak{h}=\Omega \,H\,
\Omega^{-1}$ while choosing the {\it ad hoc} ``Dyson's map''
$\Omega$ {\em non-unitary}. Thus, as long as this implies that
 \be
 \mathfrak{h}^\dagger
 = [\Omega^{-1}]^\dagger
 \,H^\dagger\, \Omega^\dagger
 =\Omega \,H\, \Omega^{-1}
 \ee
i.e.,
 \be
 H^\dagger\, \Omega^\dagger\,\Omega
 = \Omega^\dagger\,\Omega \,H\,,
 \ee
one has to deal with a new, effective Hamiltonian which appears {\em
manifestly non-Hermitian} in all of the generic, non-trivial
situations in which the superposition of the two Dyson's maps
remains nontrivial, $\Omega^\dagger\,\Omega=\Theta\neq I$.

From the point of view of mathematics the latter unexpected
observation did not in fact lead to any really serious
complications. Indeed, people (including, and listed by, the authors
of \cite{Geyer}) quickly imagined that although one has $H \neq
H^\dagger$ in the ``effective'', bosonic Hilbert space ${\cal
H}^{(F)}$ with the ``usual'' inner product (here, although the
superscript $^{(F)}$ stands for ``friendly'', this acronym may be
also re-read as ``false'', due to the apparent violation of the
unitarity of the evolution generated by the non-self-adjoint
generator $H$), one may simply redefine the inner product using the
``metric operator'' $\Theta$. In this way one arrives at another,
{\em physical bosonic Hilbert space} ${\cal H}^{(S)}$. The
superscript $^{(S)}$ stands here for ``standard'' because inside the
new, {\em unitarily non-equivalent} bosonic Hilbert space ${\cal
H}^{(S)}$, the evolution generated by the {\em same} generator {\it
alias} Hamiltonian $H$ appears now, in full agreement with the
standard textbooks on quantum theory, {\em unitary}.

We may summarize that the introduction of the ``standard''
Hilbert-space-metric operator $\Theta=\Omega^\dagger\Omega \neq I$
enables us to replace the original, correct but
computation-unfriendly physical representation space ${\cal
H}^{(P)}$ by its unitarily equivalent amendment ${\cal H}^{(S)}$
(which may be called, deservedly, ``sophisticated'' \cite{SIGMA}).
The whole idea may be given the form of a diagram
 \ben
  \vspace{-1cm}
  \ba
    \begin{array}{|c|}
 \hline
 \vspace{-0.3cm}\\
  {\rm  \fbox{\bf {P}}}\\
 \ {\rm textbook \  level\ quantum\ theory}\ \\
 {\rm \bf prohibitively\ complicated}\ {\rm \ Hamiltonian}\ \mathfrak{h}\\
 {\rm generating \ unitary\  time\ evolution}\ \\
  \ \ \ \ {\rm  calculations\ =\  practically\ impossible\ } \ \\
    \ \ \ \ {\rm  \ {\bf physics\ of \ traditional \ textbooks} } \ \\
 \hline
 \ea
 \\
 \stackrel{{\bf  simplification}\ \Omega^{-1}}{}
 \ \ \ \
  \swarrow\ \  \  \ \ \ \ \ \
 \ \ \ \ \ \ \ \
  \ \  \  \ \ \ \ \ \
 \ \ \ \ \  \searrow \nwarrow\
 \stackrel{\bf   equivalence}{}\\
 \begin{array}{|c|}
 \hline
 \vspace{-0.35cm}\\
  {\rm  \fbox{\bf {F}}}\\
   {\rm  physical\ meaning \ = \ lost}  \\
  {\rm   \ kets}\ |\psi\rangle \ {\rm represented}\\
  {\rm in \ \bf false\ } {\rm Hilbert\ space,} \\
    \ {\rm {\bf  feasible\   calculations} } \\
  \hline
 \ea
 \stackrel{ {\bf  hermitization}  }{ \longrightarrow }
 \begin{array}{|c|}
 \hline
 \vspace{-0.35cm}\\
  {\rm  \fbox{\bf {S}}}\\
  {\rm metric\ \Theta\ in \  inner\ product:}  \\
    H=H^\ddagger=\Theta^{-1}H^\dagger\Theta  \\
  {\rm \bf standard\ interpretation }
 \\
      {\rm \bf sophisticated\ } \Theta=\Omega^\dagger\Omega \\
 \hline
 \ea
\\
\\
\\
\ea
 \een
We see that for formal reasons it is recommended that the $P
\leftrightarrow S$ unitary equivalence is realized in two steps.
Firstly, the kets $|\psi\pkt \in {\cal H}^{(P)}$ (notice their
``curved'' denotation as proposed in \cite{SIGMA}) are interpreted
as Dyson-map images $|\psi\pkt=\Omega|\psi\kt$ of their ``bosonic''
representants in an auxiliary, unphysical, ``false'',
$F-$superscripted Hilbert space, $|\psi\kt \in {\cal H}^{(F)}$.
Secondly, a new inner product is introduced, in ${\cal H}^{(F)}$,
just to define, formally, another, viz., the second physical Hilbert
space ${\cal H}^{(S)}$.

The main advantage of the resulting representation of a given
quantum system via a triplet of Hilbert spaces may be seen in the
underlying implicit assumption of the {\em thorough simplification}
$\mathfrak{h} \to H$ of the Hamiltonian paid by an {\em affordable
complication} of the Hermitian conjugation in the physical Hilbert
space. Indeed, we have to replace the traditional ``transposition
plus complexification'' maps (viz., $\mathfrak{h} \to
\mathfrak{h}^\dagger$ and  $H \to H^\dagger$) in ${\cal H}^{(P)}$
and ${\cal H}^{(F)}$, respectively, by their more sophisticated,
metric-dependent analogue $H \to
H^\ddagger=\Theta^{-1}H^\dagger\Theta$ in ${\cal H}^{(S)}$.

\subsection{Non-Hermitian differential-operator Hamiltonians}

Naturally, the key assumption of the decisive simplicity of the
``new'' Hamiltonian $H$ has been successfully verified not only in
the above-mentioned realistic context of the models of nuclei but
also, say, for the first-quantized Klein-Gordon equation
\cite{alikg}. An exceptional methodical role has been played by the
Buslaev's and Grecchi's ``wrong-sign'' anharmonic oscillator
\cite{BG} where {\em both} of the Hamiltonian-operator
representatives $H=H^{(BG)}$ and $\mathfrak{h}=\mathfrak{h}^{(BG)}$
preserved the elementary differential-operator form containing just
a local interaction potential, viz.,
 \be
 H^{(BG)} =\frac{1}{2}\left (
 -\frac{d^2}{dy^2} + \frac{j^2-1}{4(y-i\varepsilon)^2}
 \right )
 -g^2(y-i\varepsilon)^4\,,\ \ \ y \in \mathbb{R}
 \label{BGline}
 \ee
and
 \be
 \mathfrak{h}^{(BG)} = -\frac{d^2}{dx^2}
 +(1-gx)^2x^2-\frac{1}{2}(2gx-1)\,,\ \ \ x \in \mathbb{R}\,
 \label{BGpline}
 \ee
(see also \cite{Jonesbg}; incidentally, even the special $g=0$ case
of this model proved worth a rediscovery \cite{ptho}).

In the case of many other, generic complex local potentials entering
the differential operators
 \be
 H =
 -\frac{d^2}{dy^2} + V(y)\,,\ \ \ y \in \mathbb{R}
 \label{modeof}
 \ee
many authors revealed that the spectrum may still remain real,
discrete and bounded from below
\cite{DB,BB,DDT}. In other words, every
demonstration that a mathematically tractable operator
(\ref{modeof}) possesses such a  spectrum opens the possibility of
assigning, to it, the status of an observable (i.e., e.g., of a
Hamiltonian) of a hypothetical quantum system, i.e., the status of a
self-adjoint operator in an {\em ad hoc} physical Hilbert space
${\cal H}^{(S)}$.

Several theoretical as well as practical challenges emerge. Even in
the context of pure mathematics one quickly reveals that any -- even
approximate -- construction of the metric $\Theta$ is by far not
easy \cite{171}. The difficulties of mathematical nature are
accompanied by their phenomenological parallels. The most important
one lies in the generic loss of the locality of the operator
$\Theta$ which is reflected by the loss of the observability of the
coordinate. This may have a destructive impact upon the traditional
``kinetic plus potential energy'' tractability of
Hamiltonians~(\ref{modeof}). Their ``point-particle'' interpretation
may get lost \cite{171be}, certain ``no-go'' theorems emerge in the
context of scattering \cite{Jones}, etc.

Amazing as it may seem, all of these difficulties may prove more
than compensated by the perspective of innovations (in this context,
Refs.~\cite{Carl,ali} offer a useful reading). An encouragement of
realistic quantum model-building may be sought also in the
flexibility of the phenomenologically motivated choice of
non-trivial metrics. For illustration one may recall
Ref.~\cite{alikg} where the old problem of proper quantum-mechanical
interpretation of Klein-Gordon equation (describing, e.g., the
physics of pionic atoms) has been resolved via the use of $\Theta
\neq I$.

In our present paper we intend to point out that the quantum theory
in its three-Hilbert-space-representation (THSR) form may find one
of its fairly interesting new illustrations and applications in the
apparently traditional context of Eq.~(\ref{modeof}) where one would
merely add the new assumption that the potential $V(y)$ itself is
singular and, hence, not well defined along the real line of $y$.
This is, in fact, the situation which was assumed in our recent
paper \cite{sextsing} where we felt inspired by the toy model
(\ref{BGline}) of Ref.~\cite{BG} and where we studied
Schr\"{o}dinger equation
 \be
 \left [-\frac{d^2}{ds^2}+U(s)
 \right ]\,\phi_n(s)=E_n\,\phi_n(s)
 \,,\ \ \ \ \ \
 U(s)\ \equiv \ V[x(s)]=
 (s - {\rm i}\varepsilon)^2 + \frac{g^2}{(s - {\rm
 i}\varepsilon)^6}\,
 \label{SEdum}
 \ee
in which $s \in \mathbb{R}$, i.e., in which one had to assume that
$y \notin \mathbb{R}$ in the context of Eq.~(\ref{modeof}).

In Ref.~\cite{sextsing} we emphasized that the obligatory starting
point of the applications of the THSR formalism should be seen in
the demonstration of the reality of the spectrum. For the present
review-paper purposes these considerations will be briefly
summarized in section \ref{reco}. In subsequent section
\ref{perturex} we shall then extend the class of potentials as well
as the scope of the method. We shall complement the large$-g$
approximate constructions of section \ref{reco} by a systematic
higher-order perturbation-expansion technique. This will enable us
to study the singular models at finite couplings $g \ll \infty$.

Being well aware of the fact that our perturbation series may have
(and probably do have \cite{Bjerrum}) a vanishing radius of
convergence, our perturbation-series description of the systems with
finite couplings will be complemented and paralleled (i.e., tested,
in section \ref{ripa}) by an independent numerical reconstruction of
the spectrum using the so called Riccati-Pad\'{e} method \cite{RPM}.
We shall reveal that for a number of specific choices of strongly
singular potentials, extremely small error bars may be reached by
our perturbation-series estimates.

A successful confirmation of the applicability of the two
independent methods to a new class of quantum models in a fairly
nonstandard domain of their implementation will be obtained. Still,
our main message will lie elsewhere. Our numerically obtained values
of the low lying bound-state energies will be re-read as not
possessing, within error bars, any imaginary components. In other
words, a ``spectrum-reality'' confirmation will be declared covering
the dynamics beyond the currently published area of not too singular
potentials.

Our present observations will finally be discussed in section
\ref{summary}. Naturally, our study of singular potentials still
leaves multiple open questions unanswered. It must be emphasized,
nevertheless, that whenever available, the better, rigorous
mathematical proofs of the reality of spectra of non-Hermitian
Hamiltonians (\ref{modeof}) appeared almost prohibitively
complicated even for regular potentials \cite{DDT,DDTb}.

%
%

\section{The key problem: the reality of the spectrum\label{regup}}

Let us reemphasize that in our present class of models with complex
``coordinates'' $y$ and ``potentials'' $V(y)$ the conventional
self-adjoint nature of the Hamiltonian is certainly lost in the
``false'' space ${\cal H}^{(F)}$. Still, whenever one proves that
the resulting spectrum of energies is real and discrete and bounded
from below, the way is open towards the reconstruction of the
appropriate Hilbert space ${\cal H}^{(S)}$ in which our Hamiltonian
becomes self-adjoint. Let us now recall, for illustration purposes,
a few most elementary examples.

\subsection{Radial regular examples}

In 2001, Dorey, Dunning and Tateo \cite{DDT} considered various
complex potentials $V(y)$ in Schr\"{o}dinger equations and they
attracted the reader's attention, in particular, to the weakly
singular centrifugal-like components $V^{(DDT)}(y) =\ell(\ell+1)/
y^{2}$ of the interactions. As long as the threshold behavior of the
general wavefunctions then remained easily tractable in closed form,
 $$
 \psi(y) \sim c_1 y^{\ell+1} + c_2 y^{-\ell}\,,\ \ \ \ |y| \ll 1,
 $$
these authors were allowed to ignore the singularity whenever they
restricted the ``coordinate'' $y$ to the lower half of the complex
plane,  ${\rm Im}\ y = -\varepsilon <0$.

Now we see the difference. For the real $y$ one would only be
allowed to work on a half-axis, with $y \in (0,\infty)$. For the
complex $y=s-{\rm i}\varepsilon$ the boundary conditions must
necessarily be changed -- while one omits the redundant boundary
condition in the origin $y=0$, a new constraint emerges as the left
asymptotic boundary condition enters the scene at $s \to -\infty$.
Naturally, not only physics (i.e., the interpretation of
measurements) but also mathematics (i.e., typically, spectra -- see,
for example, \cite{ptho}) get changed.

\subsection{A wrong-sign quartic example}

Buslaev and Grecchi \cite{BG} were probably the first who
demonstrated, constructively, that the Dyson's isospectral mapping
$\Omega$ between Hamiltonians may cause a truly thorough change of
the Hilbert space. In {\em loc. cit.} they constructed the mapping
between Hamiltonians (\ref{BGline}) and (\ref{BGpline}) and showed
that in the former operator the change of the parameter
$\varepsilon$ does not change the spectrum at all. In the light of
the analyticity properties of the potentials such an observation is
not too surprising. At the same time, in the context of
Ref.~\cite{sextsing} the same freedom of the choice of $\varepsilon$
(which, in fact, meant the freedom of a parallel shift of the
complex line of $y$ in Schr\"{o}dinger equation) proved to be of
fundamental importance.

Any generalization of the Buslaev's and Grecchi's results to a less
exceptional potential acquires, as a rule, the form of an
approximate construction. In this spirit, the new and particularly
challenging strongly singular  and ${\cal PT}-$symmetrically
regularized inverse-sextic-repulsion model of Eq. (\ref{SEdum}) will
be now considered as a methodical guide as well as one of the most
natural candidates for an apparently non-Hermitian (in the ``false''
space ${\cal H}^{(F)}$) but (in the ``standard'' space ${\cal
H}^{(S)}$) still unitarily evolving quantum model.



\section{Strongly repulsive potentials
\label{reco}}

The reasons of our interest in the strongly singular and
complex-shift regularized quantum models are explained in Appendix A
below. Our discovery of their mathematical appeal dates back to
Ref.~\cite{Omar} where their perturbative tractability has been
revealed and tested. Perturbation expansions were found to work
there for a broad complex-valued subfamily of {\em regular}
potentials. In this sense, it will only be necessary to demonstrate
here that the presence of the strongly repulsive barriers need not
obstruct the applicability of the same large$-g$
perturbation-expansion techniques.

\subsection{Schr\"{o}dinger equations at special values of
$\varepsilon$}

The main weak point of the rather universal large$-g$
perturbation-expansion technique as explained, e.g., in
Ref.~\cite{Bjerrum} is that its convergence to exact results cannot
be guaranteed in general. One only has to use the formalism as a
source of suitable asymptotic series and approximants.
%
%
In this sense, such a perturbation recipe will still satisfy our
present needs sufficiently well.

The essence of the formalism lies in several assumptions. Firstly,
%
%
%
%
%
%
%
one must require that in the regime of large couplings $g$ the
potential develops a pronounced minimum with a negligible imaginary
component. Thus, the first derivative of the potential function must
vanish at a certain complex value of the coordinate $x=R_m$. One
writes $V'(R_m)=0$ and treats such a formula as an elementary
algebraic equation determining, implicitly, the unknown eligible
complex minima $R_m$. For our model of Eq.~(\ref{SEdum}), in
particular, the latter equation reads
 $
 2R_m^8 = {6g^2}
 $
and leads to the eight well-separated closed-form candidates for the
minimum,
 \ben
 R_m=R\,e^{ {\rm i}\,\pi\,(m-1)/4}\,,\ \ \ \
 R=|3^{1/8}g^{1/4}|\gg 1\,,\ \ \ \
 m =
 1,2,\ldots,8\,.
 \een
In order to obtain an efficient approximation recipe we need to
require the positivity of the second derivative of our potential at
its stationary points. Such a necessary condition is restrictive and
not too easily satisfied. Fortunately, the verification yields the
same real quantity at all $m$ for our toy model,
 \ben
 V''(R_m)= 2+42\,\frac{g^2}{R^8_m}= 2+42\,\frac{g^2}{3g^2}=16\,.
 \een
In a systematic analysis we reveal that the uppermost root $R_3$
lies on the cut. A hardly solvable regularized double-well
Schr\"{o}dinger equation is also obtained if one selects the complex
line of its integration as intersecting the pairs of stationary
points $R_2$ and $R_4$ or $R_1$ and $R_5$ or $R_8$ and $R_6$.

\begin{figure}[h]                     
\begin{center}                         
\epsfig{file=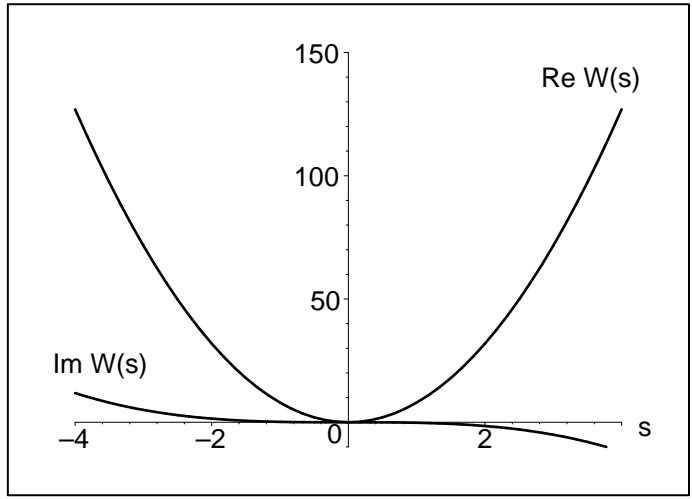,angle=270,width=0.6\textwidth}
\end{center}                         
\vspace{-2mm} \caption{The real and imaginary parts of
the upwards-shifted potential function
of Eq.~(\ref{potuj}), with $W(s)=4R^2/3+V[x(s)]$ and $\varepsilon=R=100$.
\label{firm}\label{firmonej}}
\end{figure}


%

The last and only eligible candidate for a ``useful'' stationary
point is the purely imaginary root $R_7$ which is, in this sense,
unique. We come to the conclusion that once we let the complex line
$x(s)$ cross the complex point $R_7$ (it is easy to show that this
means that we choose $\varepsilon =R$) we have satisfied all
requirements. Consequently, along the line $x(s)$ the real part of
our potential is really characterized by a pronounced minimum while
its imaginary component becomes negligible in the regime of large
parameters $R$ (cf. Fig.~\ref{firm}). Hence, we may try to apply the
recipe of Ref.~\cite{Omar}. Let us now describe the results in full
detail.

\subsection{Harmonic-oscillator approximation}

Let us recall our potential,
 \be
  V(x)= x^2 + \frac{R^8}{3x^6}\,.
  \label{potuj}
  \ee
At the large real $R$ and at the small complex shifts ${\rm
i}\varepsilon= s - x(s)$ the shape and $s-$dependence of function
(\ref{potuj}) remains dominated by its singularity. For the large
$\varepsilon \sim R$ this domination gets suppressed. The complex
function $V[x(s)]$ acquires a deep minimum at $x=-{\rm i}R$ (cf.
Fig.~\ref{firmonej}) so that we may try to Taylor-expand the
potential near the point $R_7 \equiv -{\rm i}R$,
 \ben
 V[x(s)]
 = -\frac{4}{3}{R}^{2}+8{s}^{2}-{\rm i}{\frac {56}{3R}}{s}^{3}-
 \frac{42}{R^2}{s}^{4}+{\rm i}{\frac
 {84}{{R}^{3}}}{s}^{5}+\frac{154}{{R}^{4}}{s}^{6}-
 \een
 \be
 -{\rm i}{ \frac
 {264}{{R}^{5}}}{s}^{7}-\frac{429}{{R}^{6}}{s}^{8}+{\rm i} \frac {
 2002}{3{R}^{7}}{s}^{9}+
 \frac{1001}{{R}^{8}}{s}^{10}-\ldots\,.
 \label{taylor}
 \ee
The decrease of the higher-order terms appears so quick that the
harmonic-oscillator term is dominant and that the radius of
convergence of the series remains large, equal to $R$. The most
important observation is that along the line of the integration of
the equation, the imaginary components of the potential become
entirely negligible.


%


The polynomial truncations of series (\ref{taylor}) remind us of the
popular complex power-law interaction models (cf. Ref.~\cite{Carl}).
Once we restrict our attention just to the first two terms of series
Eq.~(\ref{taylor}) we even arrive at the exactly solvable model of
the usual, real harmonic oscillator.
%
%
Its low-lying spectrum of bound states is well known yielding the
fairly reliable approximation
 \be
 E_n=-\frac{4{R}^{2}}{3}+(2n+1)\sqrt{8} + {\cal O}\left ( \frac{1}{R}
 \right )\,,
 \ \ \ \ \ n = 0, 1, \ldots\,.
 \ee
All of the higher-order contributions  lead just to asymptotically
vanishing corrections to the energies. Thus, the first few orders of
perturbation theory lead to the low-lying energy levels which are
all equidistant, real and negative.

We may summarize that the main message delivered by this section is
that under the assumption that the real coupling constant $g$ is
kept very large, the low lying spectrum of energies can be found, in
spite of the manifest non-Hermiticity of Hamiltonian $H_1^{(IS)}$,
real. One may expect that the Hamiltonian may be reinterpreted again
as self-adjoint in an {\em ad hoc} space ${\cal H}^{(S)}$.

\section{Systematic perturbation expansions \label{perturex}}

Our considerations of preceding section were restricted to large $g
\gg 1$. After the publication of the first results of this type via
arXiv \cite{sextsing}  our present team of authors has been
established to analyze the possibilities of an extension of these
observations to the less fictitious dynamical regime of smaller,
finite couplings $g<\infty$ in the same potential as well as  to
some other, more general singular potentials, say, of the two-term
form
 \begin{equation}
 V(x)= -(\mathrm{i}x)^{2+\alpha} - \frac{g^2}{(\mathrm{i}
 x)^{6+\beta}}\,.  \label{defpot}
\end{equation}
Each of these functions may enter the ordinary differential
Schr\"{o}dinger equation subject to the same replacement $x \to
x(s)=s-{\rm i}\varepsilon$  as above.

\subsection{The case of real potentials as a methodical
guide\label{predem}}

Hamiltonian operators of the usual self-adjoint form (where $V(x)$
is real, non-singular and confining) admit the approximate
determination of the low lying spectrum of bound states via a
localization of the real minimum $x_0 \in \mathbb{R}$ of the
potential, i.e., via a determination of the root $x_0$ of a
transcendental algebraic equation $ V^{\prime }(x_{0})=0$ under
constraint $ V^{\prime \prime}(x_{0})>0$ \cite{Bjerrum}. In such a
case one can also amend the approximation using perturbation theory.
After a change of variables $x=x_{0}+\beta s$ where  $-\infty
<s<\infty $ and where $\beta $ is an arbitrary auxiliary real
scaling factor, one may expand $V(x_{0}+\beta s)$ in Taylor series
near $s=0$,
\begin{equation}
V(x_{0}+\beta s)=\sum_{j=0}^{\infty }V_{j}\beta ^{j}s^{j}\,.
\end{equation}
In the new Hamiltonian operator
\begin{equation}
H=\beta ^{-2}\left( -\frac{d^{2}}{ds^{2}}+V_{2}\beta
^{4}s^{2}+\sum_{j=3}^{\infty }V_{j}\beta ^{j+2}s^{j}\right)
\end{equation}
we choose $\beta =V_{2}^{-1/4}$ so that
\begin{equation}
H=\sqrt{V_{2}}\left( -\frac{d^{2}}{ds^{2}}+s^{2}+\sum_{j=1}^{\infty
}\frac{ V_{j+2}}{V_{2}}\beta ^{j}s^{j+2}\right) .
\end{equation}
Now one decides to apply perturbation theory to Schr\"{o}dinger
equation
\begin{equation}
\frac{1}{\sqrt{V_{2}}}H\psi =\epsilon \psi
\end{equation}
and one obtains the usual perturbation series for the eigenvalues
\begin{equation}
\epsilon =\sum_{j=0}^{\infty }\epsilon _{j}\beta ^{j}\,.
\end{equation}
By construction one has $\epsilon _{0}=\epsilon _{0}(v)=2v+1$,
$v=0,1,\ldots $. Since the transformation $(\beta ,s)\rightarrow
(-\beta ,-s)$ leaves the Hamiltonian invariant we may conclude that
$\epsilon _{2j+1}=0$ at all $j=0,1,\ldots $.

In the above context the main mathematical idea lying beyond the
considerations of Ref.~\cite{sextsing} (cf. also section \ref{reco}
above) is that all of the main components of the above construction
may remain applicable even if one leaves the real axis of $x$ and if
one performs an analytic continuation of potential $V(x)$ into the
complex plane of $x$. The only news is that in the complex case, the
potential need not have the required real-harmonic-oscillator minima
at all. {\em Vice versa}, the existence of these very specific
minima (as shown, constructively, above) should be perceived as a
very specific feature of certain ``privileged'' potentials. In other
words, not all potentials would prove tractable by the present
method.

\subsection{Perturbation expansions near a complex minimum
$x_0$\label{priu}}

Let us now return to our family of potentials of Eq.~(\ref{defpot})
and let us try to determine all of its {\em complex }  stationary
points from the vanishing-derivative condition $V^{\prime}=0$. For
methodical purposes we may just contemplate the small exponents
$\alpha$ and $\beta$. We find out that the eligible negative
imaginary stationary point $x_0$ as obtained at $\alpha=\beta=0$
(i.e., the complex coordinate $x_{0}=R_7=-\mathrm{i}T$ in the
notation of Ref.~\cite{sextsing}) merely moves to an amended
$\alpha\neq 0 \neq \beta$ candidate for the minimum
$x_{min}=-\mathrm{i}T$ where the real quantity $T$ is such that
(say, at positive $\alpha$ and $\beta$)
\[
T^{8+\alpha+\beta}=g^2\frac{6+\beta}{2+\alpha}\,.
\]
At this stationary point we also evaluate the second derivative
exactly,
\[
V^{\prime\prime}(x_{min})=(2+\alpha) T^\alpha \,(8+\alpha+\beta)\,.\
\]
This number is real and positive so that the assumptions of the
applicability of the method of paragraph \ref{predem} are satisfied.
The imaginary part of the potential remains negligible and the
leading-order harmonic-oscillator approximation of
Ref.~\cite{sextsing} will keep working.

We are now prepared to Taylor-expand the potential. Besides the
zero-order term
\[
-{\frac {{T}^{2+{\alpha}} \left( 8+{\alpha}+{\beta} \right)
}{6+{\beta}}}
\]
and besides the vanishing first order term we evaluate easily also
the second order term
\[
1/2\, \left( 2\,{\beta}+{\alpha}\,{\beta}+16+10\,{\alpha}+{\alpha}
^{2} \right) {T}^{{\alpha}}\,.
\]
The not too exciting news are coming with the third-order correction
\[
1/6\,i \left( 2+{\alpha} \right) {T}^{-1+{\alpha}} \left(
{\alpha}+{\ \alpha} ^{2}-56-15\,{\beta}-{\beta}^{2} \right)
\]
where, incidentally, the last bracket factorizes, $\left(
8+{\alpha}+{\beta} \right) \left( {\alpha}-7-{\beta} \right)$.
Finally, the fourth-order part of the potential reads
\[
-1/24\, \left( 2\,{\beta}^{3}+{\alpha}\,{\beta}^{3}+24\,{\alpha}
\,{\beta} ^{2}+48\,{\beta}^{2}+\right .
\]
\[
\left . +191\,{\alpha}\,{\beta}+382\,{\beta} +502\,{\alpha}-{
\alpha}^{2}+2\,{\alpha}^{3}+{\alpha}^{4}+1008 \right)
{T}^{-2+{\alpha}}
\]
and may be simplified as well. Indeed, the bracket factorizes again,
\[
\left( 2+{\alpha} \right) \left( 8+{\alpha}+{\beta} \right) \left(
{\alpha} ^{2}-8\,{\alpha}-{\alpha}\,{\beta}+63+16\,{\beta} +
{\beta}^{2} \right)\,.
\]
The long factor still factorizes over
 \[
\alpha_\pm =4+1/2\,{\beta}\pm 1/2\,\sqrt
{-188-48\,{\beta}-3\,{\beta}^{2}}
 \]
where the discriminant will vanish at the real roots $\beta=-8\pm
2/\sqrt{3}$. One may also add that the alternative factorization
over
\[
\beta_\pm =1/2\,{\alpha}-8\pm 1/2\,\sqrt {-3\,{\alpha}^{2}+4}
\]
seems simpler.

Once one moves to the higher orders of perturbation series, an
explicit display of formulae would become clumsy and
counterproductive. Still, it is possible to store the formulae in
the computer and use them just for an evaluation of numerical
predictions.

The computer-supported analysis and numerical tests of these results
will be shown to enable us to conclude that the discussion as given
in  Ref.~\cite{sextsing} remains applicable also to the more
complicated potentials. The inclusion of the new parameters $\alpha$
and $\beta$ does not change the overall qualitative picture of the
spectra. In what follows, we shall need just a routine procedure for
obtaining the perturbation series approximation up to the $M-$th
term, yielding the approximate energy values $E^{(M)}_n$ of the
$n-$th bound state in broad intervals of finite couplings $g$.

\subsection{Sample choices of the integer values of $\alpha$ and
$\beta$\label{vzorkyex}}

We will extend the results of  Ref.~\cite{sextsing} in two ways.
Firstly we shall perform explicit calculations while restricting our
attention to certain special integer values of $\alpha$ and $\beta$,
considering the family of singular potentials
\begin{equation}
V(x)=x^{2m}+\frac{\lambda}{x^{2n}}\,,\ \ \ \
\lambda=\frac{mR^{2(m+n)}}{n}\,,  \ \ \ \ \,m,n=1,3,\ldots \,.
\label{integy}
\end{equation}
In the spirit of preceding paragraph \ref{priu} we redefined the
coupling, $g \to R(g)$, and choose $m$ and $n $ as positive
integers.

Formally, we may now apply perturbation theory up to arbitrarily
large order. In this manner we obtain numerical results which may be
compared with some other, nonperturbatively obtained values of bound
state energies (cf. section \ref{ripa} below). Technically, the
construction of the perturbation series will be facilitated by the
friendlier notation of Eq.~(\ref{integy}).

First of all we notice that if $V(-x)=V(x)$ and if $a$ and $s$ are
real, then $U(s)=V(ia+s)$ has the property
$U(-s)^{*}=V(-ia-s)=V(ia+s)=U(s)$ called, usually, ${\cal
PT}-$symmetry. Once we consider just the family of spiked
oscillators (\ref{integy}) where $\,\lambda >0$ and where $m$ and
$n$ are positive integers, the first minimality condition $V^{\prime
}(x_{0})=0$ yields
\begin{equation}
x_{0}=\left( \frac{\lambda n}{m}\right) ^{1/[2(m+n)]}e^{{\rm i}\pi
k/(m+n)},\,\ \ k=0,1,\ldots ,2(m+n)-1.
\end{equation}
The root $x_{0}$ remains purely imaginary if we require that $m+n$
is even.  This choice may simplify the discussion and it will be
preferred in what follows. It also implies that for the second
derivative we have
\begin{equation}
V^{\prime \prime }(x_{0})=4m(m+n)x_{0}^{2(m-1)}.
\end{equation}
We see that $V^{\prime \prime }(x_{0})>0$ only if $m$ is odd. Thus,
whenever we want to keep our present discussion fully analogous to
the one of  Ref.~\cite{sextsing}, the value of $n$ should be also
chosen odd.

\section{The numerical determination of the
energies\label{ripa}}

\subsection{Quadratically convergent numerical method}

For our present purposes it is important that the quickly convergent
Riccati-Pad\'{e} method (RPM, \cite{RPM}) of the numerical
determination of the eigenvalues of Hamiltonians is well adapted
also to the treatment of the present, spatially asymmetric and
complex potentials.

The key idea of the method is that one considers a correct wave
function together with its logarithmic derivative
\begin{equation}
f(x)=-\frac{\psi ^{\prime }(x)}{\psi (x)}
\end{equation}
which, obviously, satisfies Riccati equation
\begin{equation}
f^{\prime }(x)-f(x)^{2}+V(x)-E=0\,.  \label{eq:Riccati_asymm}
\end{equation}
Once we Taylor-expand
\begin{equation}
f(x)=\sum_{j=0}^{\infty }f_{j}\left( x-x_{0}\right) ^{j}
\label{eq:f_series}
\end{equation}
and substitute (\ref{eq:f_series}) into (\ref{eq:Riccati_asymm}) we
obtain the coefficients $f_{j}$ in terms of the two unknowns $E$ and
$f_{0}=-\psi ^{\prime }(x_{0})/\psi (x_{0})$. Next, splitting the
sequence of coefficients into two subsequences,
\begin{equation}
f_{e,j}=f_{2j},\;\ \ \ f_{o,j}=f_{2j+1},\;j=0,1,\ldots
\end{equation}
we obtain both $E$ and $f_{0}$ as the roots of the system of the two
coupled nonlinear algebraic equations
\begin{eqnarray}
H_{De}^{d}(E,f_{0}) &=&\left| f_{e,i+j+d-1}\right| _{i,j=1}^{D}=0\,,
\nonumber
\\
H_{Do}^{d}(E,f_{0}) &=&\left| f_{o,i+j+d-1}\right| _{i,j=1}^{D}=0\,
\end{eqnarray}
(cf. Refs.~\cite{RPM} for more details).

\subsection{Centrifugal-like spikes,
$n=1$}

Before one applies the RPM numerical technique to singular models it
seems useful to test the approach on a minimally singular example
with $n=1$. Conveniently, we may then introduce the slightly
modified, regularized logarithmic derivative
\begin{equation}
f(x)=\frac{\sigma}{x}-\frac{\psi ^{\prime }(x)}{\psi (x)}\,.
\label{eq:f(x),n=1}
\end{equation}
With $\sigma=\left(1\pm \sqrt{4R^{4}+1}\right)/2$ this choice
exactly removes the pole of $\psi ^{\prime }(x)/\psi (x)$ at the
origin. In this case $f(x)$ satisfies the modified Riccati equation
\begin{equation}
f^{\prime }+\frac{2\sigma}{x}=f(x)^{2}+E-x^{2m}\,.
\label{eq:Riccati,n=1}
\end{equation}
With the ansatz
\begin{equation}
f(x)=x\sum_{j=0}^{\infty }f_{j}(E)x^{2j} \label{eq:f_expansion_n=1}
\end{equation}
we obtain the accurate eigenvalues in the form of the roots of the
modified Hankel determinants $H_{D}^{d}(E)=\left| f_{i+j+d+1}\right|
_{i,j=1}^{D}$. For sufficiently large $D$  we may, typically, choose
$d=0$. In this  manner the application of the RPM
philosophy becomes  more efficient than in the generic case.

\subsection{The test of the large-order perturbation
results\label{ripabe}}


As long as we restricted our quantitative analysis to the subfamily
(\ref{integy}) of potentials (\ref{defpot}), we may proceed in the
closest parallel with section \ref{reco}. In particular, we may
again put $x_{0}=-iR$, i.e., choose the same, very special and
``user-friendly'' distance $\varepsilon=R$ from the real line.

Beyond the leading-order-approximation framework as accepted in
Ref.~\cite{sextsing} we may now compare the exact numerical RPM
predictions with those given by the semi-analytic perturbation
expansions. In our present generalized, two-parametric models the
function $U(x)=V(x-iR)$ still does not exhibit any singularities and
it still possesses just a single global and pronounced minimum at
$x_0$. Asymptotically this function is smooth and behaves as the
power $x^{2m}$ when $|x|\rightarrow \infty $.

This implies that {\em a priori} we may expect, in general, a good
performance of perturbation theory. In order to test this
expectation we decided to carry out the respective calculations for
several lowest eigenvalues and for the decreasing sequence of
couplings $R=20,10,2,1.5$ and $1$.

For illustration the ``exact'' RPM results for ground states are
displayed first, in Table~\ref{tab:ene}. The Table shows the lowest
eigenvalue obtained by means of the RPM technique for several values
of exponents $m$ and $n$ as discussed in paragraph \ref{vzorkyex}
above.

\begin{table}[]
\caption{The lowest RPM (i.e., exact) eigenvalue for several spiked
oscillators} \label{tab:ene}
\begin{tabular}{||c|c|c||}
\hline\hline \multicolumn{1}{||c|}{$R$} &
\multicolumn{1}{c|}{$m=n=1$} & \multicolumn{1}{c||}{$m=1,\ n=3$} \\
\hline
20 & -798.00062499975585957 & -530.50539390089880261 \\
10 & -198.00249998437519531 & -130.50687623309973953 \\
5 & -48.009999000199950014 & -30.513112355071711530 \\
2 & -6.0622577482985496524 & -2.5774188753856708289 \\
1.5 & -2.6097722286464436550 & -0.35218352259563351294 \\
1 & -0.23606797749978969641 & 0.848803366333102053806
 \\ \hline & $m=3,\ n=1$ & $m=n=3$
\\ \hline
20 & -255998040.41035854784 & -127997600.00562501215 \\
10 & -3999510.1106623371002 & -1999400.0225007778344 \\
5 & -62377.559943373182961 & -31100.090049835809532 \\
2 & -236.61574750381748921 & -104.5752232235887004 \\
1.5 & -34.899091771582975476 & -10.358453053216073663 \\
1 &  -12.250254250322260411  & 1.52979838806531408256 \\

\hline\hline
\end{tabular}
\end{table}

Our subsequent tests of the performance of the perturbation series
(i.e., of the precision of the $N-$th order perturbative
ground-state approximants $E^{[N]}$) were based on the direct
comparison with the RPM-based numerical results (of any prescribed
precision). A sufficiently representative sample of these
comparisons is summarized in Fig.~\ref{fig:HPM}. We used there a
characteristic ``intermediate'' coupling constant $R=2$ and the same
values of the exponents $m$ and $n$ as in Table~\ref {tab:ene}
above.

The picture demonstrates, in general, the good performance  of the
partial sums $E^{[N]}$ of order $N$. We see that the logarithmic
error of the perturbation series $\log \left| \left(
E^{[N]}-E^{RPM}\right) /E^{RPM}\right|$ decreases, in most cases, in
a sufficiently long interval of the perturbation order $N$. Still,
the individual forms of the interaction carry certain specific
characteristics.

\begin{figure}[]
\begin{center}
\includegraphics[width=14cm]{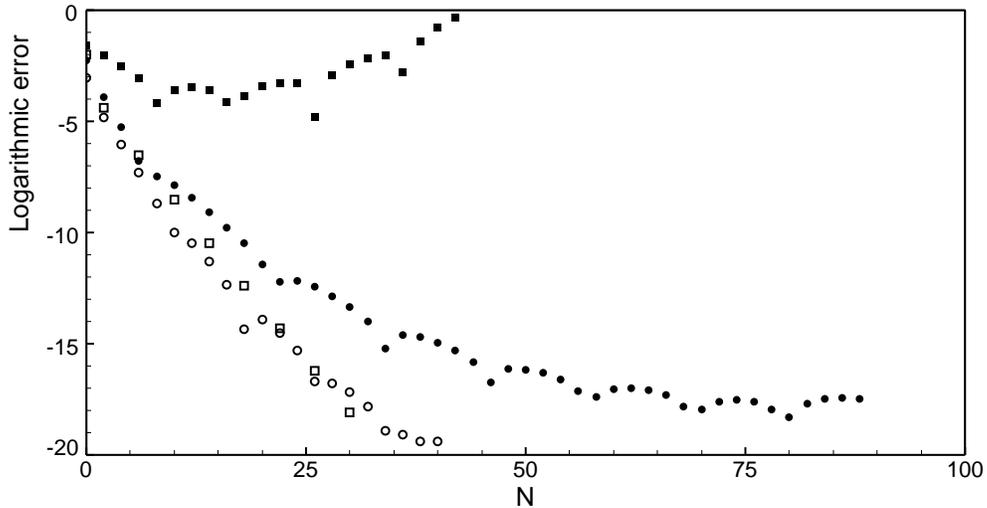}
\end{center}
\caption{Logarithmic error $\log \left| \left(
E^{[N]}-E^{RPM}\right) /E^{RPM}\right| $ for the ground state with
$R=2$ for the oscillators given by $m=n=1$ (squares), $m=1,\ n=3$
(filled squares), $m=3,\ n=1$ (circles) and $ m=n=3$ (filled
circles)}. \label{fig:HPM}
\end{figure}

\subsubsection{Potential $V(x)=x^{2}+{\lambda}/{x^{2}}\,$.}

At $m=n=1$ our approach leads to simplifications because $\epsilon
_{4j}=0$, $j=1,2,\ldots $. As expected the rate of convergence of
perturbation series decreases with $R$. It even appears to converge
at the coupling as small as $R=1$. On the other hand, the RPM
numerical technique yields $20$ accurate digits with $D=2$ ($d=0$)
disregarding the value of $R$. Incidentally, the $m=n=1$ eigenvalue
problem (i.e., the ${\cal PT}$ symmetric harmonic oscillator of
Ref.~\cite{ptho}) is exactly solvable so that we may be sure that
both our methods yield approximate results that converge towards the
exact ground-state energy $E_{00}=2-\sqrt{4R^{4}+1}$.

\subsubsection{Potential $V(x)=x^{2}+{\lambda}/{x^{6}}\,$ of section
\ref{reco},
i.e., $m=1,\,n=3$.}

In this case perturbation expansion starts to oscillate,
considerably, at a not yet too small value of $R=2$. The RPM appears
to converge for all $R$ but the rate of its convergence decreases
with the decreasse of $R$ rather quickly. One could say that among
our quadruplet of toy  models the oldest model of
Ref.~\cite{sextsing} seems least open towards perturbation-expansion
amendments.

\subsubsection{Potential $V(x)=x^{6}+{\lambda}/{x^{2}}\,$, i.e.,
$m=3,\,n=1$.}

Perturbation expansions start to oscillate at smaller $R=1.5$ and
they are, at all $R$, more stable than in the preceding example.
Incidentally, also the rate of the convergence of the purely
numerical RPM results is perceivably higher than in the preceding
case.

\subsubsection{Potential $V(x)=x^{6}+{\lambda}/{x^{6}}\,$, i.e.,
$m=n=3$.}

In picture \ref{fig:HPM} we see that in spite of certain growth of
complexity of the underlying formulae the numerical rate of
convergence of the perturbation-theory approximations remains very
satisfactory and does not exhibit oscillations even at the very high
orders. Empirically one may reveal a certain apparent regularity (or
periodicity in a monotonic decrease of perturbation-series errors)
but no immediate explanation of this phenomenon seems available to
us at present.

\section{Summary  \label{summary}}

The recent extension of quantum theory to Hamiltonians
$H=-d^2/dx^2+V(x) \neq H^\dagger$ where the coordinate is not
observable, $x \notin \mathbb{R}$, was reviewed here and tested in
an innovative context of the strongly singular potentials, say, of
the form $V(x)=(ix)^{const}\,x^2 + g^2/[(ix)^{const'}x^2]$. Three
standard (viz., large$-g$, perturbation-expansion and numerical
Riccatti-Pad\'{e}) construction techniques were shown applicable to
such a class of models. All of these methods were shown to yield
mutually compatible results supporting the hypothesis of reality of
the energies of the low-lying bound states, i.e., of a potentially
self-adjoint nature of the Hamiltonian in a properly chosen {\em ad
hoc} physical Hilbert space ${\cal H}^{(S)}$ with a nontrivial
metric $\Theta \neq I$.

Originally, a similar, purely empirical observation of the possible
reality of energies of bound states in a repulsive singular ${\cal
PT}$ symmetric potential was made, in Ref. \cite{sextsing}, in a
very restricted dynamical regime of very large repulsion strength $g
\to \infty$. In our present continuation of this study we proceeded
to the sub-asymptotic dynamical regime of finite couplings
$g<\infty$ {\em alias} $R < \infty$.

What is shared here with  Ref.~\cite{sextsing} is the explicit
clarification of the highly nontrivial physical non-equivalence and
of a deep phenomenological contrast between the traditional,
real-half-axis choice of the radial coordinate $r$ and the
innovative, ${\cal PT}-$symmetry-inspired complex (i.e., anomalous,
unobservable) choice of the line of integration of the seemingly not
too much different Schr\"{o}dinger equation. In fact, the two
complementary parts of an entirely new field of research in quantum
theory are encountered. This fact  may be perceived as a welcome and
encouraging extension of possibilities of the model-building in
quantum mechanics.

One should also mention that in the future analyses of singular
models our deeper understanding of the possible underlying physics
will require a further extension of the construction, in particular,
towards some (i.e., at least, leading-order-form) metrics which
would define the above-mentioned ``Hermitizing'' inner products in
Hilbert space ${\cal H}^{(S)}$. What one could find encouraging also
in this context is, for the present particular choice of interaction
models, the large-coupling negligibility of the imaginary part of
the potentials near their minima at $x_0$.

It is necessary to add that the subject itself is by far not
exhausted. First of all, our present, RPM-based demonstration of the
absence of the imaginary parts in energies is more or less purely
numerical. Secondly, the perturbation method we tested remains
restricted just to the low-lying part of the spectrum of energies.
Even its compatibility with independent numerical RPM results does
not offer a rigorous proof of course. {\em A priori}, within our
present methodical framework one still cannot exclude the
possibility of the presence of some exponentially small imaginary
components in the energies.

This being said, our non-rigorous numerical results may still be
declared important because their existence strengthens the
intuitively sound persuasion that the time evolution of the
underlying quantum systems cannot deviate from unitarity too much.
In other words, one may expect that on a pragmatic and approximative
level it is possible to consider our user-friendly new Hamiltonians
as operators with a good chance of being perceived also as ${\cal
PT}-$symmetric and  self-adjoint with respect to a suitable,
nontrivial physical inner product in the underlying Hilbert space of
admissible quantum states.

On the background provided by our previous paper \cite{sextsing} let
us add that what is also provided by the present new models, methods
and calculations is a long-expected extension of the leading-order
approximations towards a systematic formalism of full-fledged
perturbation theory. Certain semi-analytic features (i.e.,
Taylor-series nature) of our parallel RPM computations might be,
perhaps, also re-classified as bringing an independent new insight
into the structure of the spectra and, in particular, of the wave
functions.

\newpage

\section*{Appendix A: Two-parametric family of
regularized singular interactions: phenomenological
aspects.\label{modely}}

One of the key merits of the bound-state Schr\"{o}dinger equations
of the ordinary differential form
\begin{equation}
\left [-\frac{d^2}{dx^2}+V(x) \right ]\,\psi_n(x)=E_n\,\psi_n(x) \,,
\ \ \ \ n = 0, 1, \ldots  \label{SEstra}
\end{equation}
is that they combine a broad phenomenological applicability and
methodical appeal with the formal friendliness of the linear
differential equations of the second order. This has been
re-emphasized in  Ref. \cite{sextsing} where  a judiciously chosen
next-to-harmonic toy-model potential was studied in a specific
strong-repulsion dynamical regime in which $g \gg 1$. In a
continuation and generalization of this analysis (cf. section
\ref{perturex} here) one introduces a coordinate-dependent
generalization of the couplings,
\begin{equation}
V(x)= V_{\alpha,\beta,g}(x)=x^2\,\mu_\alpha(x) +
\frac{g^2\,\nu_\beta(x)}{x^6}\,.
  \label{function}
\end{equation}
Both of the additional non-constant functions of $x$ possess the
same one-parametric power-law forms of $\mu_\alpha(x)= (ix)^{\alpha}
$, $\alpha \geq 0$ and $\nu_\beta(x)=1/(ix)^{\beta}$, $\beta \geq
0$.

This made the shape of the potential more flexible. Moreover, one
may return to the original potential via an elementary limiting
transition $\alpha \to 0$ and $\beta \to 0$. Far from this limit, on
the contrary, the shape of the function(s) $\mu_\alpha(x)$ and
$\nu_\beta(x)$ may be adapted more easily to phenomenological needs.

The list of formal reasons for our choice further incorporates also
the quasi-solvable nature of similar forces (cf. the fifth item in
Table Nr. 1 of Ref.~\cite{QESsing}), i.e., the feature which was
made popular in monograph \cite{Ushveridze}) or the tractability of
at least some of the related eigenvalue problems using continued
fractions \cite{jasing} or a specific simplicity of the asymptotic
estimates of wave functions \cite{letter}.

It makes sense to add that the studies of non-Hermitian but
real-spectra-exhibiting quantum models may be perceived as one of
the most dynamical branches of development of quantum theory after
1998 (see, e.g., reviews \cite{Carl,ali,Dorey}). One of the fairly
productive subbranches of these efforts was devoted to the
mathematical idea (which may be dated back to the early nineties
\cite{BG,BT}) that the spectrum of bound states may be in fact
controlled and modified by the mere \emph{ad hoc} redefinition of
the integration path of $x \in \mathcal{S} \subset \mathbb{C}$ (cf.,
e.g., \cite{BBsqw}).

A consequent further extension of the latter mathematical idea
(related closely to the presence of the strong singularities in
$V(x)$ but getting us already beyond the limits of our present
considerations) may be based on the question of what happens when
the localization of the underlying integration path $\mathcal{S}$ is
allowed to leave the plain complex plane (endowed, possibly, with a
cut oriented upwards). In this direction it has been proposed
\cite{tobog1} that in the definition of the integration path
$\mathcal{S}$ one may and should try to replace the (cut) complex
plane $\mathbb{C}$ by a more general Riemann surface $\mathcal{R}$.
In the latter scenario (cf. also~\cite{shendr,toborev}) one treats
the general Riemann surface $\mathcal{R}$ as composed, in usual
manner, of a set of individual Riemann-sheet cut planes,
$\mathcal{R}= \bigcup_j \mathcal{R}_j$ where $\mathcal{R}_j \sim
\mathbb{C}$. Then the path $\mathcal{S}$ of integration may and
should encircle the branch-point singularities of $\mathcal{R}$,
giving rise to several alternative, non-equivalent quantum systems
living on the respective ``tobogganic'' complex curves. Thus,  every
such a system is described not only by the ordinary differential
equation but also by the topologically nontrivial tobogganic path
$\mathcal{S}$ (connecting, in general, several individual Riemann
sheets) and, in addition, by a suitable definition of inner product
in the underlying sophisticated physical Hilbert space
$\mathcal{H}^{(S)}$ (cf. \cite{shendr} for a deeper discussion of
the latter point in tobogganic context).



\newpage

\begin{thebibliography}{99}


















\bibitem{Geyer}
F. G. Scholtz, H. B. Geyer and F. J. W. Hahne, Ann. Phys.
(NY) 213 (1992) 74.


\bibitem{SIGMA}
M. Znojil, SIGMA 5 (2009) 001, arXiv: 0901.0700. 


\bibitem{Dyson}
F. J. Dyson,
Phys. Rev. 102 (1956) 1217.


\bibitem{alikg}
A. Mostafazadeh, 
Class Quantum Grav 20 (2003) 155.
%



\bibitem{BG}
V. Buslaev and V. Grecchi, 
J. Phys. A: Math. Gen. 26 (1993) 5541.

\bibitem{Jonesbg}
H. F. Jones and J. Mateo,
Phys. Rev. D 73 (2006) 085002.



\bibitem{ptho}
M. Znojil,
Phys. Lett. A 259 (1999) 220.
%
%


\bibitem{DB}
D. Bessis, private communication (1992).

%
%


\bibitem{BB}
C. M. Bender and S. Boettcher,
Phys. Rev. Lett. 80 (1998) 5243.
%
%

\bibitem{DDT}
P. Dorey, C. Dunning, R. Tateo, J. Phys. A: Math. Gen.
34 (2001) 5679.

\bibitem{171}
A. Mostafazadeh, J. Phys. A: Math. Gen. 39 (2006) 10171.

\bibitem{171be}
A. Mostafazadeh and A. Batal, J. Phys.
A: Math. Gen. 37 (2004) 11645.

\bibitem{Jones}
H. F. Jones, Phys. Rev. D 76 (2007) 125003.
%


\bibitem{Carl}
C. M. Bender, Rep. Prog. Phys. 70 (2007) 947.

\bibitem{ali}
A. Mostafazadeh, 
Int. J. Geom. Meth. Mod. Phys. 7 (2010) 1191.

\bibitem{sextsing}
M. Znojil, 
Int. J. Theor. Phys. 53 (2014) 2549.

\bibitem{Bjerrum}
N. E. J. Bjerrum-Bohr, J. Math. Phys. 41 (2000) 2515.

\bibitem{RPM}
F. M. Fern\'{a}ndez, Q. Ma and R. H. Tipping, Phys. Rev. A 39 (1989)
1605;

F. M. Fern\'{a}ndez, Q. Ma, and R. H. Tipping, Phys. Rev. A 40
(1989) 6149.

\bibitem{DDTb}
K. C. Shin, J. Math. Phys. 42 (2001) 2513.

\bibitem{Omar}
M. Znojil, F. Gemperle and O. Mustafa,
J. Phys. A: Math. Gen. 35 (2002) 5781.



\bibitem{QESsing}
M. Znojil, 
J. Phys. A: Math. Gen. 15 (1982) 2111.


\bibitem{Ushveridze}
%
%
A. G. Ushveridze, Quasi-Exactly Solvable Models in Quantum
Mechanics. (IOP, Bristol, 1994).
%
%

\bibitem{jasing}
M. Znojil,
J. Math. Phys. 31 (1990) 108.

\bibitem{letter}
M. Znojil,
Phys. Lett. A 158 (1991) 436.



\bibitem{Dorey}
P. Dorey, C. Dunning and R. Tateo,
J. Phys. A: Math. Theor. 40 (2007) R205.

\bibitem{BT}
C. M. Bender and A. Turbiner, Phys. Lett. A 173 (1993) 442.

\bibitem{BBsqw}
C. M. Bender, S. Boettcher, H. F. Jones and V. M. Savage, J. Phys.
A: Math. Gen. 32 (1999) 6771.

\bibitem{tobog1}
M. Znojil,
Phys. Lett. A 342 (2005) 36.
\bibitem{shendr}
M. Znojil, 
J. Phys. A: Math. Theor. 41 (2008) 215304.
%
%

\bibitem{toborev}
%
%
C. M. Bender, D. W. Hook, and S. P. Klevansky, J. Phys. A: Math.
Theor. 45 (2012) 444003. 



\end{thebibliography}
\end{document}